\documentclass[a4paper]{IEEEtran} 

\usepackage{amsmath}
\usepackage{amssymb}
\usepackage{amsfonts}
\usepackage{amsthm}
\usepackage{epsfig}
\usepackage{subfigure}
\usepackage{calc}
\usepackage{color}
\usepackage[all]{xy}
\usepackage{bm}
\usepackage{algorithmicx}
\usepackage{algpseudocode}
\usepackage{algorithm}
\usepackage{enumerate}
\usepackage{cite}
\usepackage{multicol}

\newtheorem{defn}{Definition}
\newtheorem{thm}{Theorem}[section]
\newtheorem{cor}[thm]{Corollary}
\newtheorem{prop}{Proposition}

\newtheorem{lem}[thm]{Lemma}
\newtheorem{conj}[thm]{Conjecture}
\newtheorem{constr}[thm]{Construction}
\newtheorem{note}{Remark}

\newcommand{\bit}{\begin{itemize}}
\newcommand{\eit}{\end{itemize}}
\newcommand{\bcor}{\begin{cor}}
\newcommand{\ecor}{\end{cor}}
\newcommand{\beq}{\begin{equation}}
\newcommand{\eeq}{\end{equation}}
\newcommand{\beqn}{\begin{equation*}}
\newcommand{\eeqn}{\end{equation*}}
\newcommand{\bea}{\begin{eqnarray}}
\newcommand{\eea}{\end{eqnarray}}
\newcommand{\bean}{\begin{eqnarray*}}
\newcommand{\eean}{\end{eqnarray*}}
\newcommand{\ben}{\begin{enumerate}}
\newcommand{\een}{\end{enumerate}}
\newcommand{\bdefn}{\begin{defn}}
\newcommand{\edefn}{\end{defn}}
\newcommand{\bnote}{\begin{note}}
\newcommand{\enote}{\end{note}}
\newcommand{\bprop}{\begin{prop}}
\newcommand{\eprop}{\end{prop}}
\newcommand{\blem}{\begin{lem}}
\newcommand{\elem}{\end{lem}}
\newcommand{\bthm}{\begin{thm}}
\newcommand{\ethm}{\end{thm}}
\newcommand{\bconj}{\begin{conj}}
\newcommand{\econj}{\end{conj}}
\newcommand{\bconstr}{\begin{constr}}
\newcommand{\econstr}{\end{constr}}
\newcommand{\bpf}{\begin{proof}}
\newcommand{\epf}{\end{proof}}

\title{Locality-Aware Hybrid Coded MapReduce for Server-Rack Architecture}
\begin{document}
\author{
\IEEEauthorblockN{Sneh Gupta and V. Lalitha\\}
\IEEEauthorblockA{SPCRC, International Institute of Information Technology, Hyderabad\\
Email: \{sneh.gupta@research.iiit.ac.in, lalitha.v@iiit.ac.in\}\\}
\vspace{-0.7cm}
}
\date{\today}
\maketitle

\begin{abstract}
MapReduce is a widely used framework for distributed computing. Data shuffling between the Map phase and Reduce phase of a job involves a large amount of data transfer across servers, which in turn accounts for increase in job completion time. Recently, Coded MapReduce has been proposed to offer savings with respect to the communication cost incurred in data shuffling. This is achieved by creating coded multicast opportunities for shuffling through repeating Map tasks at multiple servers. We consider a server-rack architecture for MapReduce and in this architecture, propose to divide the total communication cost into two:  intra-rack communication cost and cross-rack communication cost. Having noted that cross-rack data transfer operates at lower speed as compared to intra-rack data transfer, we present a scheme termed as Hybrid Coded MapReduce  which results in lower cross-rack communication than Coded MapReduce at the cost of increase in intra-rack communication. In addition, we pose the problem of assigning Map tasks to servers to maximize data locality in the framework of Hybrid Coded MapReduce as a constrained integer optimization problem. We show through simulations that data locality can be improved considerably by using the solution of optimization to assign Map tasks to servers.

%
        
\end{abstract}

\section{Introduction}

Distributed computing systems are becoming more widespread, playing an increasing role in our everyday computational tasks.
 There are two aspects to these systems: one is distributed storage of data and the other is distributed computing. An example of a platform which allows both distributed storage and computing on big data is Hadoop~\cite{Hadoop}.
The distributed storage part of Hadoop is known as the Hadoop distributed file system (HDFS). In HDFS,  any file that needs to be stored is divided into blocks and these blocks are replicated, typically $3$ times and are stored across the distributed storage network, such that no two replicas of the same block are stored in the same storage node.

The distributed computing framework of Hadoop is known as MapReduce, which allows for processing large datasets on a cluster of commodity servers. The MapReduce framework is divided into the following phases:
\begin{itemize}
\item {\bf Map Phase:} A job is divided into Map tasks and assigned to different servers. The output of each Map task is an intermediate output, which itself is given as input to the reduce phase. 
\item {\bf Shuffle Phase:} In the Shuffle phase, the intermediate map outputs are transferred among the servers to provide input to the Reduce phase. This process of exchanging intermediate map outputs on the network is referred to as {\em data shuffling}. 
\item {\bf Reduce Phase:}  In Reduce phase, the intermediate map outputs are aggregated/reduced to produce the final results of the job.
\end{itemize}

Within this framework, data shuffling often limits the performance of distributed computing applications. In a FacebookÕs Hadoop cluster, it is observed that one-third of the overall job execution time is spent on data shuffling \cite{LiAliAve}.

\vspace{-0.1in}

\subsection{Coded MapReduce}

Coded MapReduce has been introduced in \cite{LiAliAve} to reduce the communication cost incurred during data shuffling. Coded MapReduce employs the following two ideas in order to create coding opportunities for data shuffling so that the inter-server communication is lowered.
\begin{itemize}
\item {\em Repeated Map Tasks:} The map task on the same subfile is repeated on different servers.
\item {\em Combining Map task outputs and multicasting:} The output of the map tasks are in the form of $<$key, value$>$ pairs. Instead of transmitting individual Map outputs to the reducer servers, the scheme allows for transmitting composite outputs of the form $<$(key1,key2), f(value1,value2)$>$[subfile1, subfile2]. These composite key, value pairs are multicasted to multiple servers in the network. $f(.)$ is chosen to be a linear function so that if a server already has the Map output $<$key1, value1$>$[subfile1], then it can calculate $<$key2, value2$>$[subfile2] from the transmitted pair $<$(key1,key2), f(value1,value2)$>$[subfile1, subfile2].


\end{itemize}

It has been shown in \cite{LiAliAve} that the above scheme reduces the communication cost as compared to naive scheme. Also, the savings increases with increasing the replication factor of the map tasks.

\vspace{-0.1in}
\subsection{Our Contribution} \label{sec:contrib}

In this paper, we extend the setting of Coded MapReduce to take into account two aspects in a practical distributed computing system (i) The first is intra-rack vs cross-rack communication during data shuffling (ii) the second aspect is data locality during allocation of Map tasks.

\begin{itemize}

\item {\em Intra-rack and cross-rack communication:} Distributed computing systems typically have server-rack architecture. The servers within a rack are connected via a Top of Rack switch and servers across racks are connected via a Root switch which connects several Top of Rack switches \cite{RasSha}. It is also well known that cross-rack traffic is bottlenecked by the Root switch and hence higher amount of cross-rack traffic can result in lower job completion times \cite{HoWu}. In this paper, we will consider this aspect in the context of Coded MapReduce. We propose a Hybrid Coded MapReduce scheme which offers a low cross-rack traffic at the cost of increased intra-rack traffic, analyze the performance of the scheme and compare with uncoded scheme and Coded MapReduce scheme.

\item {\em Data Locality:} A map task is said to be a {\it local} task, if the map task is assigned to a server containing the corresponding subfile (required by the task). A map task is said to be {\it remote} task if the map task is assigned to a server which does not have the subfile on which it has to perform the task. A non-local or {\it remote} task needs data to be fetched across the network, leading to increased delay in job execution as well as increased network bandwidth usage~\cite{DeaGhe}. Data locality is the percentage of total number of Map tasks which are local tasks. The Coded MapReduce scheme proposed in \cite{LiAliAve} does an assignment of Map tasks to the servers without taking into account the data locality factor. In this paper, we pose the problem of assigning Map tasks to servers such that data locality is maximised in the Hybrid Coded MapReduce scheme as constrained integer optimization problem.
\end{itemize}

\vspace{-0.2in}

\subsection{Related Work}

 Algorithms for coded data shuffling for reducing communication cost and coded matrix multiplication to mitigate straggler nodes (nodes that form bottleneck of a distributed computing job) in the context of distributed machine learning have been investigated in \cite{LeeLam}. Coding for speeding up data shuffling in the TeraSort job has been studied in \cite{Li_terasort}. A generalisation of the current setting of Coded MapReduce, dealing with the tradeoff between computation and communication has been studied in \cite{LiAliAve_tradeoff}. In \cite{AttTan}, the tradeoff between storage and communication cost incurred as part of data shuffling has been discussed. Maximizing data locality in the general MapReduce setting has been considered in \cite{GuoFox}. Optimizing cross-rack communication by careful reducer placement has been proposed in \cite{HoWu}.

\section{System Model}

In this section, we will describe the basic architecture of the servers in a distributed computing system under consideration and introduce the parameters of the system. 

\begin{figure} [h]
\begin{center}
\includegraphics[width=3.5in]{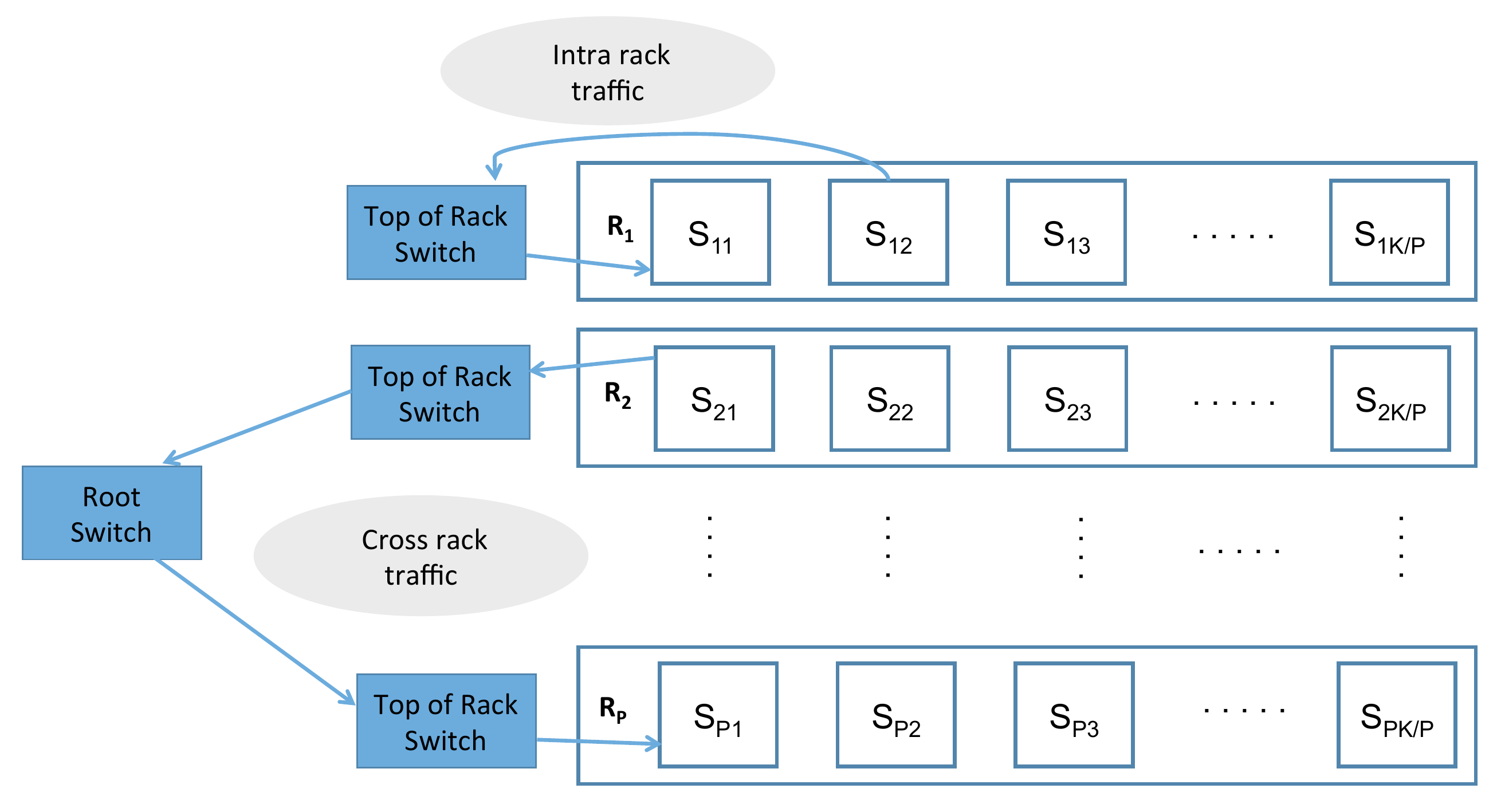}
\end{center}
\caption{Server-Rack Architecture of a Distributed Computing System.}
\label{fig:topology}
\end{figure}

Consider a distributed computing system with multiple racks and each rack consists of multiple servers as shown in Fig. \ref{fig:topology}. All the nodes within a rack are connected to a Top of Rack (ToR) switch. All the ToR switches of individual racks are connected via a root switch. We assume that the servers can multicast to other servers, i.e., a server can send a message to two or more destination (whichever servers are intended receivers for the message) servers in one shot. We assume that multicast communication between two or more servers within a rack takes place via the Top of Rack Switch (and hence high speed) and the multicast communication from a server to a set of servers which span multiple racks will involve using the root switch (and hence low speed). The total communication cost will thus be divided into two components: (i) intra-rack communication (denoted by subscript $int$) by which we will refer to number of $<$key,value$>$ pairs transferred via the ToR switch and (ii) cross-rack communication (denoted by subscript $cro$) by which we will refer to number of $<$key,value$>$ pairs transferred via the root switch. We define the following parameters of the system.

\begin{itemize} 
\item \textit{$r$ (Map Task Replication Factor)} denotes the number of times a map task is repeated.

\item \textit{$K$} denotes number of servers in the cluster.

\item \textit{$P$ } is the number of racks in the cluster.

\item \textit{$K_r$} denotes the number of servers in a rack, $K_r = \frac{K}{P}$.

\item The servers are denoted by $S_{ij}, 1 \leq i \leq P, 1 \leq j \leq \frac{K}{P}$. We will refer to the set of servers $\{S_{1i}, S_{2i}, \ldots, S_{Pi}\}$ which have the same second index as a layer of servers.
    
\item \textit{$r_f$ (File Replication)} denotes the number of replicas of a particular subfile.

\item \textit{$N$} denotes the number of subfiles of a particular MapReduce job.


\item \textit{$Q$} is the number of keys to reduce.
    
\end{itemize}


\section{Hybrid Coded MapReduce Scheme}

In this section, we will describe a Hybrid Coded MapReduce scheme and show that the scheme has much lower cross-rack communication cost $L_{cro}$ as compared to that of Coded MapReduce, at an increased cost of $L_{int}$, intra-rack communication cost. We term the scheme as Hybrid Coded MapReduce since the replication of Map tasks is done across the racks and across the servers within a rack, there is no replication of Map tasks and hence it is a hybrid of Coded MapReduce and uncoded scheme. 
%

 \begin{figure}[h!]
\begin{center}
\includegraphics[width=2.6in]{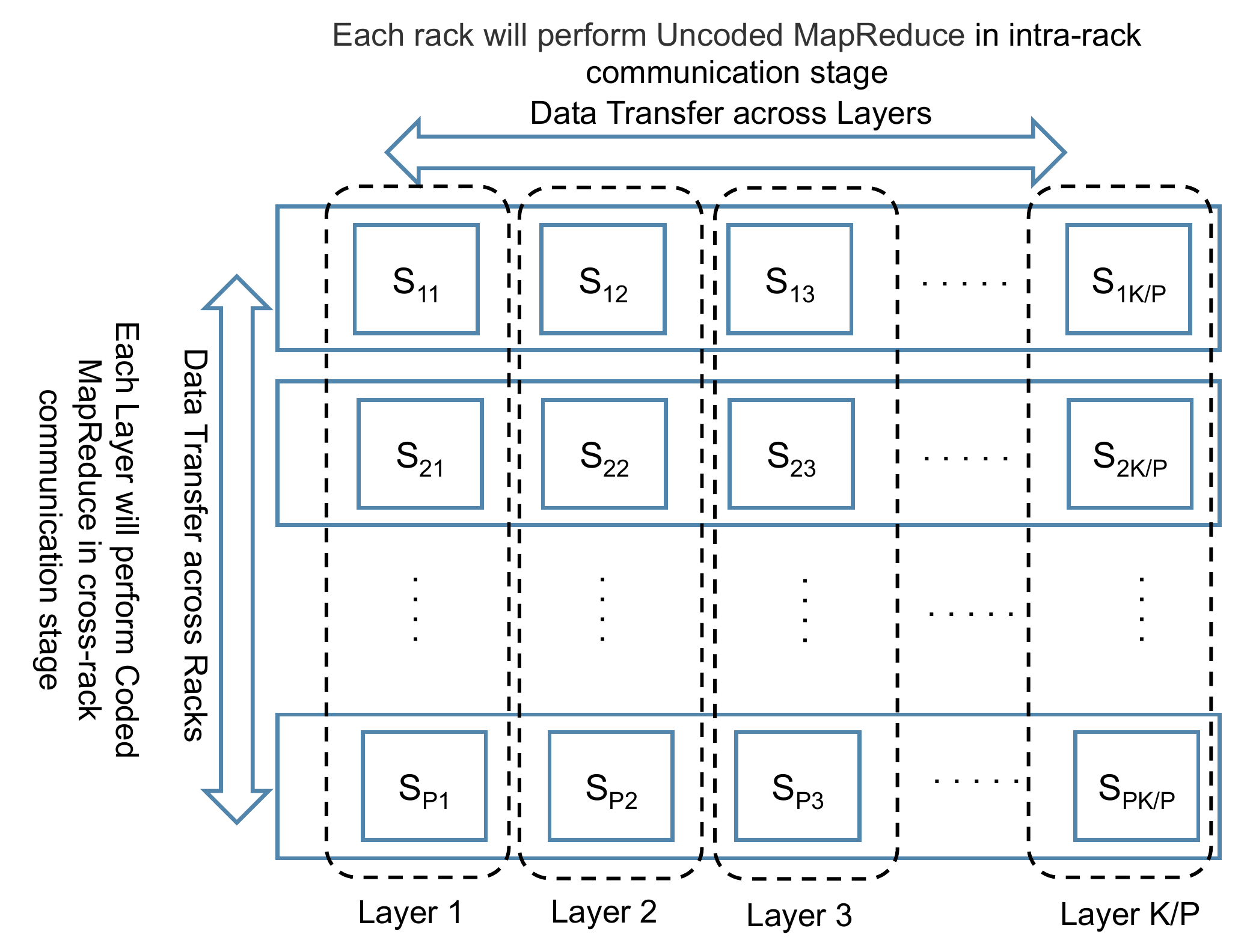}
\end{center}
\vspace{-0.1in}
\caption{Cross-Rack and Intra-Rack communication in Hybrid Coded MapReduce}
\label{fig:hybrid}
\end{figure}

The Hybrid Coded MapReduce scheme consists of the following two phases:
  \begin{enumerate}
  \item  {\em Map Task Assignment Phase:}
  \begin{enumerate}[(a)]
  \item We divide the total number of subfiles $N$ into $K_r = \frac{K}{P}$ layers (this terminology is chosen to match with that of layer of servers). The $i^{\text{th}}$ layer is denoted by $\mathcal{A}_i$. The number of subfiles in $\mathcal{A}_i$ is $\frac{NP}{K}$.
  \item We assume that ${P \choose r} | \left ( \frac{NP}{K} \right ) $  and the ratio is denoted by
   $ M = \frac{NP}{K} \frac{1}{{P \choose r}}$.
  \item Consider the set of servers $S_{1i}, \ldots, S_{Pi}$. Out of these $r$ servers can be picked in ${P \choose r}$ ways. For each choice of $r$ servers, assign a unique set of $M$ subfiles from the layer $\mathcal{A}_i$.
  \item The previous step is repeated for all the $\frac{K}{P}$ layers. This finishes the Map task assignment phase.
  \item Based on this assignment, we denote any subfile by $F_{T, w}^{(i)}$, where any $1 \leq i \leq \frac{K}{P}$ is the layer number, $T$ is a specific $r$-subset of $[1:P]$ and $1 \leq w \leq M$ is the subfile index within subfiles corresponding to  the $r$-subset $T$.
  \end{enumerate}
  
   \item {\em Data Shuffling Phase:}
  In the Hybrid Coded MapReduce scheme, the data shuffling happens in two stages. In the first stage, the cross-rack communication is performed and in the second stage, the intra-rack commmunication is carried out (See Fig. \ref{fig:hybrid}).
  $\frac{Q}{K}$ keys are reduced by each server and hence $\frac{Q}{P}$ keys are reduced by each rack. 
   \begin{enumerate}
   \item {\em Cross-rack Communication Stage:} In this stage, the data shuffling is based on the Coded MapReduce scheme and it is performed independently for each layer of servers. We will describe the process for a single layer of servers. In the cross-rack communication stage, each layer has a total of $\frac{NP}{K}$ subfiles and each server in the layer has to effectively reduce $\frac{Q}{P}$ keys (all the keys which are reduced in the rack). This is because in the next step, only intra-rack communication takes place. In the following discussion, we describe the Coded MapReduce for $\frac{NP}{K}$ subfiles, $P$ servers and $\frac{Q}{P}$ keys, with a Map task replication factor $r$.
    \begin{itemize}
    \item We choose a combination of $r + 1$ servers from the $i^{\text{th}}$ layer say $S_{1i}, \ldots, S_{(r+1)i}$.
    \item We will consider the assignment of the Map tasks. Since any set of $r$ servers have a unique set of $M$ subfiles assigned to them, there are a total of ${r+1\choose r} M = (r+1)M$ subfiles, which are completely assigned within these $r+1$ servers. 
    \item Considering the server $S_{2i}$, there are $M$ subfiles which are not assigned to $S_{2i}$, whereas they are assigned to all the other $r$ servers $S_{1i}, S_{3i} \ldots, S_{(r+1)i}$. Out of these $M$ subfiles, each of $S_{1i}, S_{3i} \ldots, S_{ (r+1)i}$ send $\frac{M}{r}$ subfiles each to $S_{2i}$ with the $\frac{Q}{P}$ keys that $S_{2i}$ requires.
    \item To consider the transmission by the server $S_{1i}$, $S_{1i}$ sends $\frac{M}{r}$ subfiles each to $S_{2i}$, $S_{3i}, \ldots, S_{(r+1)i}$ with their corresponding $\frac{Q}{P}$ keys each.
    \item The above transmissions by $S_{1i}$ can be achieved by multicasting as follows:
    \begin{eqnarray} 
   Keys & = &(G_{2,u}, G_{3,u}, \ldots, G_{r+1,u}), \nonumber \\
   Value & = & f(v_2(u,w), v_3(u,w), \ldots v_{r+1}(u,w)), \nonumber \\ 
   Subfiles & = & [F_{T_2, w}^{(i)}, \ldots, F_{T_{r+1}, w}^{(i)}],
   \label{eq:multicast}
    \end{eqnarray}
    where $1 \leq u \leq \frac{Q}{P}$, $ 1 \leq w \leq \frac{M}{r}$.  $G_{2,u}, 1 \leq u \leq \frac{Q}{P}$ denotes the set of all keys which have to be reduced by all the servers in rack 2 and so on. $v_2(u,w)$ is the value of the key $G_{2,u}$ in the subfile $F_{T_2, w}^{(i)}$ and so on. $T_z = [1: r+1]\setminus \{z\}$. $f(.)$ is a linear function of $r$ values, which satisfies the property that whenever all except one $<key, value>[subfile]$ pairs are known at a server, the unknown value can be calculated using all known values and the value $f(.)$. For example, consider the above multicast by  $S_{1i}$ as in \eqref{eq:multicast}. At server 2, all pairs $<G_{z,u}, v_z(u,w)>[F_{T_z,w}^{(i)}], 3 \leq z \leq r+1$ are known. So, based on the property satisfied by function $f(.)$, $<G_{2,u}, v_2(u,w)>[F_{T_z,w}^{(i)}]$ can be calculated at server 2. Similar inferences can be made at all the $r$ servers based on the multicast by server $S_{1i}$.
    \item The above process is repeated for all $r+1$ servers in the chosen combination. Also, it is repeated for all possible ${P \choose r+1}$ ways of choosing $r+1$ servers out of $P$ servers.
   \end{itemize}
   \item {\em Intra-rack Communication Stage:} In this stage, the data shuffling is performed within racks. Each rack has a total of $N$ subfiles, divided among $\frac{K}{P}$ servers and there is no repetition of subfiles along the rack. Each server will transmit all the keys of the $\frac{NP}{K}$ subfiles except the ones that it has to reduce. Thus, each server transmits $\frac{Q}{P} - \frac{Q}{K}$ keys for each of the $\frac{NP}{K}$ subfiles. Note that in this stage, the data transfer takes place in unicast mode and there is no need for multicast communication.
    \end{enumerate}
   This concludes the data shuffling phase of the Hybrid Coded MapReduce scheme and by the end of this phase, all the servers have all the subfiles which they require to reduce their respective keys.

    
    \end{enumerate}
    \vspace{-0.2in}
    \subsection{Analysis of Hybrid Coded MapReduce Scheme}
    In this subsection, we will use $L_{int}^{Unc}$ and $L_{cro}^{Unc}$ to refer to intra-rack communication and cross-rack communication under Uncoded MapReduce scheme. Similarly, the superscripts $Cod$ and $Hyb$ will refer to Coded MapReduce and Hybrid Coded MapReduce respectively.
    
    \begin{prop}
      Consider a system with a MapReduce job on $N$ subfiles, $K$ servers, $P$ racks, $Q$ keys such that $K | N$, $P | K$ and $K | Q$, then under uncoded MapReduce we have
    \begin{equation}
    L_{int}^{Unc}  =  Q N \left ( \frac{1}{P} - \frac{1}{K} \right ), \  L_{cro}^{Unc}  =  Q N \left ( 1 - \frac{1}{P} \right ).
    \end{equation}
    
    \end{prop}
    
    \bpf
    
    Each server is assigned $\frac{N}{K}$ subfiles. $\frac{Q}{K}$ keys have to be reduced by any server. We note that every server has to transmit $(Q - \frac{Q}{P}) \frac{N}{K}$ keys to other racks (Since $\frac{Q}{P}$ keys are reduced in the rack in which the server is present).
       Hence, the cross-rack communication cost  is given by $ L_{cro}^{Unc} = Q N (1 - \frac{1}{P})$. Further, the intra-rack communication cost is given by is $ L_{int}^{Unc} = Q N (\frac{1}{P} - \frac{1}{K})$. Finally, we have $ L_{tot}^{Unc} =   L_{int}^{Unc} + L_{cro}^{Unc} = Q N (1 - \frac{1}{K})$.
       \epf

    \begin{prop}
      Consider a system with a MapReduce job on $N$ subfiles, $K$ servers, $P$ racks, $Q$ keys and Map task replication factor $r$ such that ${K \choose r} | N $, $P | K$ and $K | Q$, then under Coded MapReduce we have
    \begin{eqnarray*}
    L_{int}^{Cod} & = & \frac{QN}{r} \left (1 - \frac{r}{K} \right ) \frac{{\frac{K}{P} \choose r + 1}} {{K \choose r+ 1}} P,  \\
    L_{cro}^{Cod} & = & \frac{QN}{r} \left (1 - \frac{r}{K} \right ) \left ( 1 - \frac{{\frac{K}{P} \choose r + 1}} {{K \choose r+ 1}}P\right ).
    \end{eqnarray*}
    
    \end{prop}

    \bpf
    
       In Coded MapReduce, we first consider the multicast within a set of $r+1$ servers. Each server within the set is multicasting $\frac{J}{r}$ subfiles (where $J = \frac{N}{{K \choose r}}$) to the other $r$ servers in the set with their respective $\frac{Q}{K}$ keys. Hence, the data transferred from one server to other servers within the set of $r+1$ servers is $\frac{Q}{K}\frac{J}{r}$.   Thus, the total data transfer for combination of $r + 1$ servers is $\frac{Q}{K}\frac{J}{r} (r+1)$. The total data transfer under Coded MapReduce scheme can be obtained as follows by considering all possible combinations of $r+1$ subsets among $K$ servers.
       \begin{equation}
    L_{tot}^{Cod} = \frac{Q}{K}\frac{J}{r} (r+1) {K \choose r+1} =  \frac{QN}{r} \left (1 - \frac{r}{K} \right ).
    \end{equation}
    A multicast communication is considered as intra-rack transfer when the combination of $r + 1$ servers is such that every server is from the same rack. Out of ${K \choose r+1}$ possibilities, $P {\frac{K}{P} \choose r+1}$ choices of sets satisfy this condition and hence, intra-rack communication cost is given by $ L_{int}^{Cod} = L_{tot}^{Cod} \frac{{\frac{K}{P} \choose r + 1}} {{K \choose r+ 1}} P$.
   Cross-rack communication cost is determined by $L_{cro}^{Cod} = L_{tot}^{Cod} - L_{int}^{Cod}$.    
    \epf

    \begin{thm}
    Consider a system with a MapReduce job on $N$ subfiles, $K$ servers, $P$ racks, $Q$ keys and Map task replication factor $r$ such that ${P \choose r} | \left ( \frac{NP}{K} \right )$, $ P | K$ and $P | Q$, then under Hybrid Coded MapReduce we have
    \begin{equation}
     L_{cro}^{Hyb}  =   \frac{QN}{r} \left (1 - \frac{r}{P} \right ), \  L_{int}^{Hyb}  =  Q N \left ( 1 - \frac{P}{K} \right ).  \\
    \end{equation}
    
    \end{thm}
    
    \bpf
    In the Hybrid Coded MapReduce, the cross-rack communication stage is equivalent to $\frac{K}{P}$ layers of Coded MapReduce on $\frac{NP}{K}$ subfiles, $P$ servers, $Q$ keys and Map task replication factor $r$. Hence, the cross-rack communication cost of Hybrid Coded MapReduce is the total data transfer cost of Coded MapReduce with the above parameters and it is given by
$    L_{cro}^{Hyb}  =  \frac{K}{P} L_{tot}^{Cod} =  \frac{K}{P} Q  \frac{NP}{K}  \frac{1}{r} ( 1 - \frac{r}{P})$.     The intra-rack communication stage is equivalent to $P$ copies of uncoded MapReduce on $N$ subfiles, $\frac{K}{P}$ servers, $\frac{Q}{P}$ keys. Hence, the intra-rack communication cost is given by $ L_{int}^{Hyb}  =  P L_{tot}^{Unc}   =  P \frac{Q}{P} N \left (1 - \frac{1}{\frac{K}{P}} \right )$.
    \epf
    
    It can be easily seen that intra-rack communication cost of Hybrid Coded MapReduce scheme is $P$ times that of Uncoded MapReduce. However, cross-rack communication cost of Hybrid Coded MapReduce is $\frac{ \frac{1}{r} - \frac{1}{P}}{1 - \frac{1}{P}}$ (approximately $\frac{1}{r}$ for large $P$) times that of Uncoded MapReduce. The comparison of communication costs of Hybrid Coded MapReduce with those of Coded MapReduce is given below.

%
%
%

\begin{cor}
The comparison of cross-rack communication cost and intra-rack communication of Hybrid Coded MapReduce against Coded MapReduce is given by
\begin{equation}
\frac{L_{cro}^{Cod}}{L_{cro}^{Hyb}}  \geq   \frac{1 - \frac{r}{K}}{1 - \frac{r}{P}} \left ( 1 - \frac{e^{r+1}}{P^r}\right ), \frac{L_{int}^{Hyb}}{L_{int}^{Cod}}  \leq   r \frac{K-P}{K-r} e^{r+1} P^r.
\end{equation}
\end{cor}
\bpf
The above inequalities follow by taking the ratios of the communication costs derived above and using the following upper and lower bounds on the binomial coefficient
$\left (\frac{n}{k} \right )^k \leq {n \choose k} < \left (\frac{ne}{k} \right )^k$. For large $P$,  $L_{cro}^{Hyb}$ is better than $L_{cro}^{Cod}$ by a factor $\frac{1 - \frac{r}{K}}{1 - \frac{r}{P}}$, at the cost of increased intra-rack communication. The ratio of intra-rack communication costs between Hybrid Coded MapReduce and Coded MapReduce is bounded above by a factor which grows polynomial in $P$. 
\epf

\vspace{-0.2in}

\section{Data Locality Optimization}

We have discussed Hybrid Coded MapReduce which provides a method to replicate Map tasks across servers. We can observe that if there is an assignment of subfiles $1, \ldots, N$ to servers under Hybrid Coded MapReduce, then any permutation of the $N$ subfiles also gives an assignment. In this section, we will consider the problem of finding the optimum permutation of subfiles such that data locality is maximised for the case when the Map replication factor $r=2$.  We will pose the problem as constrained integer optimization problem. Consider a system with $N$ subfiles, $K$ servers, $P$ racks, $Q$ keys and Map task replication factor $r=2$ such that ${P \choose 2} | \left ( \frac{NP}{K} \right )$ (the ratio is denoted by $M$), $P | K$ and $P | Q$.
To describe the optimization problem, we will index all the $K$ servers with a single index (instead of double indices which we were using till this point). The index of server $S_{ij}$ is given by $(i-1)\frac{K}{P}+j$. Now, we will introduce certain variables which will be used to indicate an assignment and data locality corresponding to the assignment.
\begin{enumerate}[(a)]
\item $X(i,j,k), 1 \leq i  \leq N, 1 \leq j \neq k \leq K$ takes value $1$ if the subfile $i$ is assigned to the server pair $(j,k)$ and $0$ otherwise. Based on the definition, we have and $X(i,j,k) = X(i,k,j)$. Also, we fix the variables $X(i,j,j)=0$.
\item $Y(j,k), 1 \leq j \neq k \leq K$ takes value $1$ if server $j$ and server $k$ have a subfile in common and $0$ otherwise. We fix the variables $Y(j,j)=0$. These variables are functions of $X(i,j,k)$ (introduced for ease of notation).
\item $C(i,j,k), 1 \leq i  \leq N, 1 \leq j,k \leq K$, is a measure of data locality when subfile $i$ is assigned to the server pair $(j,k)$. Please refer to Section \ref{sec:sim} for one possible choice of $C(i,j,k)$. 
\end{enumerate}


\begin{thm} \label{thm:data_locality}
 The assignment of subfiles (interchangeably referred to as Map tasks) to the servers under Hybrid Coded MapReduce scheme to maximise data locality is equivalent to solving the following optimization problem:
\begin{equation}
\max_{\{X(i,j,k)\}} \sum_{i=1}^N  \sum_{j=1}^K \sum_{k=1}^K  X(i,j,k) C(i,j,k),
\end{equation}
subject to the following four constraints,
\begin{enumerate}
\item {\em No common files in a rack:} $X(i,j,k) = 0$ and $Y(j,k)=0$ if $\lfloor \frac{j-1}{\frac{K}{P}} \rfloor = \lfloor \frac{k-1}{\frac{K}{P}} \rfloor$.
\item {\em Common subfiles condition:}  $\sum_{i=1}^N X(i,j,k) = M Y(j,k), \ \   1 \leq j,k \leq K$.
\item {\em Degree condition:}  $\sum_{j=1}^K Y(j,k) = P-1, \ \ 1 \leq k \leq K$.
\item {\em Transitivity:} $Y(i,j) + Y(j,k) + Y(i,k) \neq 2, \forall i,j,k$ distinct. 
\end{enumerate}
\end{thm}
\bpf
The first condition has to be satisfied by a Hybrid Coded MapReduce scheme since there is no replication within a rack. We know from the Hybrid Coded MapReduce scheme that any two servers have either none or $M$ files in common, which is given by condition (2). Also, a server has common files with $P-1$ other servers in the layer which is given by condition (3). If servers $(i,j)$ have a file in common and $(j,k)$ have a file in common, then all $i,j,k$ belong to the same layer, which forces $i,k$ to have a file in common. This means that the only possible conditions are $Y(i,j) + Y(j,k) + Y(i,k) \in \{0, 1, 3\}$. Hence, we have condition (4) satisfied by the Hybrid CodedMapReduce scheme. Conversely, it can be deduced that the solution to the optimization problem results in a valid Map task assignment for Hybrid Coded MapReduce.
\epf

%

\section{Simulations and Results} \label{sec:sim}

In this section, we will present our simulations and results for the following: (i) comparing cross-rack communication and intra-rack communication for three schemes: Uncoded, Coded and Hybrid Coded MapReduce (ii) Comparing the data locality obtained by optimization based assignment and random assignment of Map tasks.


\begin{table}[h!]
\centering
\begin{tabular}{|l|l|l|l|l|l|l|}
\hline
($K$,$P$,$Q$,$N$,$r$)  &   \multicolumn{3}{c|}{Cross-Rack} & \multicolumn{3}{c|}{Intra-Rack} \\
\hline
 & Unc  & Cod & Hyb & Unc & Cod & Hyb \\
\hline
(9,  3,18, 72,  2) & 0.864         & 0.486         & 0.216          & 0.288         & 0.018          & 0.864          \\
\hline
(16, 4, 16, 240, 2) & 2.88        & 1.632        & 0.96          & 0.72         & 0.048          & 2.88         \\
\hline
(16, 4, 16, 1680, 3) & 20.16       & 6.976        & 2.24        & 5.04        & 0.304         & 20.16       \\
\hline
(15, 3, 15, 210, 2) & 2.1      & 1.275        & 0.525          & 0.84         & 0.09          & 2.520         \\
\hline
(20, 4, 20,  380, 2) & 5.7        & 3.3        & 1.9         & 1.52        & 0.12         & 0.608         \\
\hline
(25, 5 , 25, 600, 2) & 12       & 6.75        & 4.5         & 2.4        & 1.5         & 12        \\
\hline
(25, 5, 25, 6900, 3) & 138      & 50.6       & 23        & 27.6       & 0.1         & 13.8        \\
\hline
(30, 5, 30, 870, 2) & 16.56       & 11.88       & 7.83         & 3.45        & 0.3         & 17.25        \\
\hline
(30, 6 , 30, 870, 2) & 21.75       & 12       & 8.7         & 3.48        & 0.18         & 20.88 \\
\hline
\end{tabular}
\vspace{0.1in}
\caption{Comparison of cross-rack and Intra-rack communication cost for Uncoded, Coded and Hybrid MapReduce schemes. The actual cost = Corresponding Number in the table $\times 1000$. }
\label{tab:data_shuffling}
\end{table}

\begin{enumerate}

\item For comparing the three schemes, we vary the number of racks from $3$ to $6$ (Please refer Table \ref{tab:data_shuffling}). The number of servers $K$, keys $Q$ and $N$ are picked so that the required divisibility conditions are met for all the schemes. We can note from Table \ref{tab:data_shuffling} that though the total communication cost is minimum for Coded MapReduce, the cross-rack communication cost is minimum for Hybrid Coded MapReduce. Though the intra-rack transfer is high for Hybrid Coded MapReduce, we would like to note that intra-rack transfers can happen in parallel. We also note that the savings of cross-rack communication reduce, whenever for a constant number of servers, the number of racks is increased (refer to Table \ref{tab:data_shuffling}) .
\item The formulation in Theorem \ref{thm:data_locality} holds for any measure of data locality $C(i,j,k)$. In our simulations, we calculate $C(i,j,k)$ by calculating two quantities NodeLocality and RackLocality. NodeLocality is defined as the number of servers among $\{j,k\}$ in which subfile $i$ is present and can take values $\{0,1,2\}$. RackLocality is defined in a similar way by considering the racks in which the servers $j$ and $k$ are present. The data locality measure is calculated as $C(i,j,k) = \lambda \text{NodeLocality} + (1-\lambda)\text{RackLocality}, \lambda \in (0.5,1]$. $C(i,j,j)$ is set to $0$. The results in Table \ref{tab:data_locality} indicate that as the number of servers increase, both node locality and rack locality improve considerably.

\end{enumerate}

\begin{table}[h!]
\centering
\begin{tabular}{||c|c|c|c|c|c|c|c||}
\hline
K                       & P                      & $r_f$            & N                        & \multicolumn{2}{c|}{Node Locality}                           &  \multicolumn{2}{c||}{Rack Locality}             \\ 
\hline
                       &                    &             &                         &  Ran                 & Opt             & Ran                 & Opt              \\ 
\hline
8  & 2 & 2 & 160 &25\% & 60\% & 80\% & 80\%    \\ 
\hline
8  & 2 & 3 & 100 &39\% & 76\% & 95 \% & 95\%   \\ 
\hline
9  & 3 & 2 & 144 & 17\% & 64\% & 57 \% & 86\%    \\ 
\hline
9  & 3 & 3 & 90  &33\% & 87\% & 77 \% & 98\%    \\ 
\hline
10 & 5 & 2 & 100 & 19\% & 80\% & 41\%    & 92.50\% \\ 
\hline
16 & 4 & 2 & 192 & 10\% & 64\% & 45 \% & 90\%   \\ 
\hline
16 & 4 & 3 & 192 & 19\% & 84\% & 63\%  & 99\%   \\ 
\hline
18 & 3 & 2 & 180 & 11\% & 60\% & 57\%  & 83\%   \\ 
\hline
20 & 5 & 2 & 200 &13\% & 66\% & 38 \% & 90\%    \\ 
\hline
21                       & 3                      & 2                      & 84                       & 12\%                      & 63\%                      & 56\%                       & 81\%                         \\ 
\hline
\end{tabular}
\vspace{0.1in}
\caption{Comparison of Data Locality under Random assignment of Map tasks and optimization based assignment.}
\label{tab:data_locality}
\end{table}

\vspace{-0.3in}

\section*{Acknowledgment} This work was supported partly by the Early Career Research Award (ECR/2016/000954) from Science and Engineering Research Board (SERB) to V. Lalitha. 
\vspace{-0.1in}

\bibliographystyle{IEEEtran}
	\bibliography{coded_mapreduce}

\end{document}